\documentclass[fleqn,twoside]{article}
\usepackage{espcrc2}

\usepackage{graphicx}

\newcommand{\lbl}[1]{\label{eq:#1}}
\newcommand{ \rf}[1]{(\ref{eq:#1})}

\newcommand{\be}{\begin{equation}}
\newcommand{\ee}{\end{equation}}
\newcommand{\bea}{\begin{eqnarray}}
\newcommand{\eea}{\end{eqnarray}}
\newcommand{\noi}{\noindent}
\newcommand{\nn}{\nonumber}
\newcommand{\ra}{\rightarrow}
\newcommand{\Ra}{\Rightarrow}

\newcommand{\lesssim}{ {\
\lower-1.2pt\vbox{\hbox{\rlap{$<$}\lower5pt\vbox{\hbox{$\sim$}}}}\ } 
}
\newcommand{\gtrsim}{ {\
\lower-1.2pt\vbox{\hbox{\rlap{$>$}\lower5pt\vbox{\hbox{$\sim$}}}}\ } 
}

\newcommand{\cL}{{\cal L}}

\newcommand{\cN}{{\cal N}}
\newcommand{\cO}{{\cal O}}
\newcommand{\cP}{{\cal P}}

\newcommand{\cW}{{\cal W}}

\newcommand{\Imm}{\mbox{\rm Im}}
\newcommand{\Ree}{\mbox{\rm Re}}

\newcommand{\tr}{\mbox{\rm tr}}

\newcommand{\MeV}{\mbox{\rm MeV}}
\newcommand{\GeV}{\mbox{\rm GeV}}

\newcommand{\with}{\mbox{\rm with}}

\newcommand{\annd}{\mbox{\rm and}}

\newcommand{\als}{\alpha_{\mbox{\rm {\scriptsize s}}}}

\newcommand{\GF}{G_{\mbox{\rm {\small F}}}}

\newcommand{\eff}{\mbox{\rm eff}}

\newcommand{\EM}{\mbox{\rm {\tiny EM}}}
\newcommand{\QCD}{\mbox{\rm {\footnotesize QCD}}}

\newcommand{\MHA}{\mbox{\rm {\footnotesize MHA}}}

\newcommand{\stern}{\langle\bar{\psi}\psi\rangle}

\input epsf

\newcommand{\AmS}{{\protect\the\textfont2
  A\kern-.1667em\lower.5ex\hbox{M}\kern-.125emS}}

\title{Analytic Approaches to Kaon Physics}

\author{Eduardo de Rafael\address[MCSD]{CPT, CNRS--Luminy, Marseille}
\thanks{Work partly supported by the TMR, 
EC--Contract No.
ERBFMRX-CT980169 (EURODA$\Phi$NE).}}
       
\begin{document}

\begin{abstract}
Most of the analytic approaches which are used at present to
understand the low energy hadronic interactions in Particle Physics, get
their inspiration from QCD in the limit of a large number of colors $N_c$.
I first illustrate this with the example of the
left--right correlation function which is an
excellent theoretical laboratory. Next, I
present the list of observables which have been computed using a
large--$N_c$ QCD approach. Finally, I discuss in some detail
examples which are relevant to lattice QCD, in the sense that we can
make comparisons. 
\vspace{1pc}
\end{abstract}

\maketitle

\section{INTRODUCTION}
In the Standard Model, the electroweak interactions of hadrons at
very low energies are conveniently described by an effective chiral
Lagrangian, which has as active degrees of freedom the low lying
$SU(3)$ octet of pseudoscalar particles, plus leptons and photons.
The underlying theory is a  $SU(3)_{C}\times
SU(2)_{L}\times U(1)_{W}$ gauge theory which is formulated in terms
of quarks, gluons and leptons, together with the massive gauge
fields of  the electroweak interactions and the hitherto unobserved
Higgs particle. Going from the underlying Lagrangian to the 
effective chiral Lagrangian is a typical renormalization group
problem.  It has been possible to integrate the heavy degrees of
freedom of the underlying theory, in the presence of the strong
interactions, perturbatively, thanks to the asymptotic freedom
property of the $SU(3)$--QCD sector of the theory. This brings us down
to an effective  field theory which consists of the QCD Lagrangian
with the $u$, $d$,
$s$ quarks still active, plus a string of four quark operators and
mixed quark--lepton
operators, modulated by coefficients which are functions of the
masses of the fields which have been integrated out and the scale
$\mu$ of whatever renormalization scheme has been used to
carry out this integration. We are still left 
with the evolution from this effective field theory, appropriate at
intermediate scales of the order of a few $\GeV$, down to an effective
Lagrangian description in terms of the  low--lying pseudoscalar particles
which are the Goldstone modes associated with the spontaneous symmetry
breaking of chiral--$SU(3)$ in the light quark sector. The dynamical
description of this evolution involves genuinely
non--perturbative phenomena and it is mostly studied using the
techniques of Lattice QCD.  In this talk, I shall review the 
progress which has been made in approaching this last step using
analytic methods; in particular when the problem is formulated
within the context of QCD in the limit of a large number of
colours $N_c$.

The suggestion to keep $N_c$ in QCD as a free
parameter was first made by G.~'t Hooft~\cite{THFT74} as a possible
way to approach the study of non--perturbative aspects of QCD. The
limit $N_c\ra\infty$ is taken with the product $\alpha_{\mbox{\rm
{\footnotesize s}}} N_c$ kept fixed. In
spite of the efforts of many  illustrious theorists who have worked
on the subject, QCD in the large--$N_c$ limit ($\QCD_{\infty}$)
still remains unsolved; but many interesting properties have been
proved, which suggest that, indeed, the theory in this limit has
the bulk of the non--perturbative behaviour of full QCD.
In particular, it has been shown that, if confinement persists in
this limit, there is spontaneous chiral symmetry
breaking~\cite{CW80}. It should be stressed that 
$\QCD_{\infty}$ is not just a ``wild extrapolation'' from
$N_c=3$ to $N_c=\infty$. In fact, $N_c$ is really used as a label
to select specific topologies among Feynman diagrams. The 
topology which corresponds to the highest power in the $N_c$--label 
is the one which selects {\it planar} diagrams only; and the claim
is that this class 
provides already a good approximation to the full theory.

The spectrum of $\QCD_{\infty}$ consists of an infinite number of
narrow stable meson states which are flavour nonets~\cite{W79}.
This spectrum looks {\it a priori} rather
different to the one in the real world: the examples of the vector and
axial--vector spectral functions measured in $e^+ e^- \ra$ hadrons
and in the hadronic
$\tau$--decay which are shown below, have indeed a richer
structure than just a sum of narrow states. There are, however, many
instances  where one can show that the observables one is
interested in, are given by smooth integrals of {\it specific}
hadronic spectral functions, though unfortunately in most cases,
the {\it specific} hadronic spectral functions in question are not
accessible from experiments. Typical examples of such observables, are
the coupling constants of the effective chiral Lagrangian of QCD at
low energies, as well as the coupling constants of the effective
chiral Lagrangian of the electroweak interactions of pseudoscalar
particles in the Standard Model. What is needed for the evaluation
of these coupling constants is not so much
the detailed point--by--point knowledge of the relevant hadronic
spectrum, but rather a good approximation consistent with the
asymptotic properties of QCD, both at short-- and long--distances.
It is in this sense that the simple
$\QCD_{\infty}$--spectrum becomes useful. It provides a simple
parameterization of the physical hadronic spectrum, based on first
principles. 

\section{THE CHIRAL LAGRANGIAN}
The strong and electroweak interactions of the Goldstone modes at
very low energies are described by an effective Lagrangian which has
terms with an increasing number of derivatives (and quark masses if
explicit chiral symmetry breaking is taken into account.) Typical terms
of the chiral Lagrangian are
\bea\lbl{chiral}
 \cL_{\eff}  & =  & \underbrace{\frac{1}4F_{0}^2\
\tr\left(D_{\mu}U
D^{\mu}U^{\dagger}\right)}_{ 
\pi\pi\ra\pi\pi\,,\quad K\ra\pi e\nu
} \\
 & & +\underbrace{L_{10}
\tr\left(
U^{\dagger}F_{R\mu\nu}
UF_{L}^{\mu\nu}\right)}_{ \pi\ra
e\nu\gamma }+\cdots \\ 
&  &  
+\underbrace{e^{2}C
\tr\left(
Q_{R}UQ_{L}U^{\dagger}\right)}_{
-e^2 C\frac{2}{F_{0}^{2}}\left(\pi^{+}\pi^{-}+K^{+}K^{-}
\right)}+\cdots \lbl{c} 
\eea
\be\lbl{g8}
-\underbrace{\frac{\GF}{\sqrt{2}}V_{ud}V^{*}_{us} 
\ g_{\underline{8}}F_{0}^{4}\left(D_{\mu}U
D^{\mu}U^{\dagger}\right)_{23}}_{
{K\ra\pi\pi\,, \quad K\ra\pi\pi\pi}}+\cdots\,,
\ee
where $U$ is a $3\times 3$ unitary matrix in flavour space which
collects the Goldstone fields and which under chiral rotations
transforms as $U\ra V_{R}U V_{L}^{\dagger}$; $D_{\mu}U$ denotes the
covariant derivative in the presence of external
vector and axial--vector sources. The first line is the lowest order
term in the sector of the strong interactions~\cite{We79}, $F_0$ is
the pion--decay coupling constant in the chiral limit where the
light quark masses $u$, $d$, $s$ are neglected ($F_0\simeq 90~\MeV$);
the second line shows one of the couplings at
$\cO(p^4)$~\cite{GL84,GL85}; the third line shows the lowest order
term which appears when  photons and $Z's$ are integrated out ($Q_L=Q_R=
\mbox{\rm diag.}[(2/3,-1/3,-1/3]$), in the presence of the strong
interactions ; the fourth line shows one of the lowest order terms in the
sector of the weak interactions. The typical physical processes to which
each term contributes are indicated under the braces. Each term is
modulated by a coupling constant:
$F_{0}^{2}$, $L_{10}$,... $C$...$g_{\underline{8}}$..., which encodes the
underlying dynamics responsible for the appearance of the
corresponding effective term. The evaluation of these couplings from the
underlying theory is the question we are interested in. The coupling 
$g_{\underline{8}}$ for
example, governs the strength of the dominant
$\Delta I=1/2$ transitions for
$K$--decays to leading order in the chiral expansion. 

\subsection{Two crucial observations}  

There are two
crucial observations to be made concerning the relation of these low
energy constants to the underlying theory.

i) The low--energy constants of
the Strong Lagrangian, like
$F_{0}^{2}$ and 
$L_{10}$,  are the
coefficients of the \underline{\it Taylor expansion}
of appropriate QCD Green's Functions. For example, with
$\Pi_{LR}(Q^2)$  the correlation function of a
left--current with a right--current in the chiral limit:
\bea\lbl{LR}
\int \!d^4x\ e^{iq\cdot x}\langle 0\vert
T\!\left(\bar{u}_{L}\gamma^{\mu}d_{L}(x)
\bar{u}_{R}\gamma^{\nu}d_{R}(0)^{\dagger}\right)\vert
0\rangle & & \nn \\
  = \frac{1}{2i}(q^{\mu}q^{\nu}-g^{\mu\nu}q^2)\Pi_{LR}(Q^2)\,, & &
\eea
the Taylor expansion  
\be\lbl{chpt}
-Q^2\Pi_{LR}(Q^2)
\vert_{Q^2\ra
0}
=F_{0}^{2}-4L_{10}\ Q^2\!+
\cdots\,,
\ee
defines the constants $F_{0}^{2}$ and $L_{10}$. 

ii) By contrast, the low--energy constants 
of the Electroweak Lagrangian, like e.g.
$C$ and $g_{8}$, are  
\underline{\it integrals} of appropriate QCD Green's Functions. For
example~\cite{KPdeR98},
\be\lbl{pimd}
C\!=\!\frac{3}{32\pi^2}\int_{0}^{\infty}\! dQ^2 \frac{M_{Z}^2}{Q^2+M_Z^2}
\left(-Q^2\Pi_{LR}(Q^2)\right)\,.
\ee
Their evaluation appears to be,
{\it a priori}, quite a formidable task because they require the
knowledge of Green's functions at all values of the euclidean
momenta; i.e. they require a precise {\it matching} of the {\it
short--distance} and the {\it long--distance} contributions to the
underlying Green's functions.

These two observations are generic in the case of the Standard
Model, independently of the $1/N_c$--expansion. The
large--$N_c$ approximation helps, however, because it restricts the
{\it analytic structure} of the Green's functions in general, and
$\Pi_{LR}(Q^2)$ in particular, to be {\it meromorphic functions}: they
only have poles as singularities; e.g., in $\QCD_{\infty}$,
\be\lbl{largeN}
\Pi_{LR}(Q^2)\!=\!\!\sum_{V}\!\frac{f_{V}^{2}M_{V}^2}{Q^2+M_{V}^2}-
\!\sum_{A}\!\frac{f_{A}^{2}M_{A}^2}{Q^2+M_{A}^2}\!-
\!\frac{F_{0}^{2}}{Q^2}\!\,, 
\ee
where the sums are extended to an infinite number of
states. In practice, however, in the case of Green's functions which
are {\it order parameters} of spontaneous chiral symmetry breaking (like
$\Pi_{LR}$ in the chiral limit), these sums will be
restricted to a finite number of states.

\subsection{Sum Rules}

There are two types of important restrictions on Green's
functions like $\Pi_{LR}(Q^2)$. One type follows from the fact
that, as already stated above, the Taylor expansion at low euclidean
momenta must match the low energy constants of the strong chiral
Lagrangian. This  results in a series of {\it\underline{long--distance
sum rules}} like e.g.
\be
\sum_{V}f_{V}^2- \sum_{A}f_{A}^2=-4L_{10}\,.
\ee

Another type of constraints follows
from the {\it \underline{short--distance  properties}} of the
underlying Green's functions. The behaviour at large euclidean momenta
of the Green's functions which govern the low energy constants of the
chiral Lagrangian can be obtained from the operator
product expansion (OPE) of local currents in QCD. In
the large--$N_c$ limit, this results in a series of algebraic sum
rules~\cite{KdeR98} which restrict the coupling constants and masses
of the hadronic poles. In the case of the $LR$--correlation
function in Eq.~\rf{largeN} one has e.g.,
\be\lbl{1wsr}
\sum f_{V}^2M_{V}^2-\sum f_{A}^2M_{A}^2 - F_{0}^{2}=0\,,
\ee
\be\lbl{2wsr}
\sum f_{V}^2M_{V}^4-\sum f_{A}^2M_{A}^4  =0 \,,  
\ee
\be\lbl{kdersr}
\sum f_{V}^2M_{V}^6-\sum f_{A}^2 M_{A}^6 \simeq
-4\pi\alpha_{s}\stern^2\,. 
\ee
The sum rules in Eqs.~\rf{1wsr} and \rf{2wsr} are the celebrated
Weinberg sum rules, which follow from the fact that in the chiral
limit,  there are no
$\cO(1/Q^2)$ terms and no $\cO(1/Q^4)$ terms  in the OPE of
$\Pi_{LR}(Q^2)$. The third sum rule in
Eq.~\rf{kdersr}~\cite{KdeR98}, follows from the matching between the
$\cO(1/Q^6)$--terms in the $\QCD_{\infty}$ expression in Eq.~\rf{largeN}
and  the corresponding one in the OPE (evaluated at leading
$\cO(\als N_c^2)$~\cite{SVZ79}). In principle there are an infinite number
of sum rules in
$\QCD_{\infty}$ which relate the {\it masses} and {\it couplings} of
the $\QCD_{\infty}$ narrow states to the {\it local order
parameters} which appear in the OPE and to the {\it non--local order
parameters} which govern the chiral expansion in the strong
sector.     

\section{THE $\pi^{+}-\pi^{0}$ MASS  DIFFERENCE AS A THEORETICAL
LABORATORY}

The physical effect of the coupling $C$ in Eq.~\rf{c} is to give a
mass, mostly of electromagnetic origin, to the charged pions:
\be\lbl{pimdem}
m_{\pi^{\pm}}^2\big\vert_{\EM}=\frac{\alpha}{\pi}\frac{3}{4F_{0}^2}
\int_{0}^{\infty} dQ^2\left[-Q^2\Pi_{LR}(Q^2) \right]\,.
\ee
The first evaluation of this integral was done
by F.~Low and collaborators in 1967~Ê\cite{Letal67}. What I am going
to do next is, simply, to put their {\it phenomenological}
calculation within the context of $\QCD_{\infty}$. 

As already stated, the function $\Pi_{LR}(Q^2)$ in $\QCD_{\infty}$
is a {\it meromorphic} function.  That means that within a
finite radius in the complex $Q^2$--plane it only has a {\it finite
number of poles}. The natural question which arises is:  {\it what
is the minimal number of poles  required to satisfy the OPE
constraints?}  The answer to that follows from a well known theorem
in analysis~\cite{Tit39}, which when applied to our case, states
that the function
\be\lbl{delta}
-Q^2\Pi_{LR}(Q^2)\equiv\Delta[z]\quad\with\quad
z=\frac{Q^2}{M_{V}^2}\,,
\ee
and $M_{V}$ the mass of the lowest narrow state, has the property that
\be
\cN-\cP=\frac{1}{2\pi i}\oint\frac{\Delta'[z]}{\Delta[z]}dz\,,
\ee
where $\cN$ is the number of zeros and $\cP$ is the number of poles
inside the integration contour in the complex $z$--plane (a zero and/or a
pole of order m is counted m times.) For a circle of radius
sufficiently large, we simply have that
$\cN-\cP=-p\,,$
where $p$ denotes the asymptotic power fall--off in $1/z$ of the
$\Delta[z]$ function. Identifying the power $p$ with the {\it
leading power} predicted by
the OPE, and from the fact that 
$\cN\ge 0$, there follows that $\cP\ge p_{\mbox{\rm{\tiny
OPE}}}$. In our case $p_{\mbox{\rm{\tiny OPE}}}=2\Ra \cP\ge 2$  and the
{\it\underline {minimal hadronic approximation}} ($\MHA$) compatible
with the OPE requires two poles~\footnote{Notice that with the definition 
of $\Delta[z]$ in Eq.~\rf{delta} the pion pole is removed.}: one vector
state and one axial--vector state. The $\MHA$ to the $\QCD_{\infty}$
expression in Eq.~\rf{largeN} is then the simple function
\be\lbl{MHA}
-Q^2\Pi_{LR}(Q^2)=F_{0}^{2}\frac{M_{V}^2
M_{A}^2}{(Q^2+M_{V}^2)(Q^2+M_{A}^2)}\,.
\ee
Inserting this function in Eq.~\rf{pimdem} gives a prediction for the
electromagnetic
$\pi^{+}-\pi^{0}\equiv\Delta m_{\pi}$ mass difference, with the
result~\footnote{This is the result for
$F_{0}\!=\!(87\pm 3.5)\,\MeV$, $M_{V}\!=\!(748\pm 29)\,\MeV$ and
$g_{A}\!=\!\frac{M_{V}^2}{M_{A}^2}\!=\!0.50\pm 0.06$. These values follow
from an overall fit to predictions of the low energy
constants in the MHA to $\QCD_{\infty}$~\cite{PPdeR98,GP00}.}
\be\lbl{mpimha}
\Delta m_{\pi}=(4.9\pm 0.4)\,\MeV\,, 
\ee
to be compared with the experimental value~\cite{PDG00}
\be
\Delta m_{\pi}=(4.5936\pm 0.0005)\,\MeV\,. 
\ee

\subsection{The real world and the
$\MHA$ to  $\QCD_{\infty}$ }

\begin{figure}
\caption{\it The Spectral Function
$\frac{1}{\pi}\Imm\Pi_{LR}(t)$, obtained from the ALEPH
data~\cite{ALEPH} compared to the $\MHA$ to $\QCD_{\infty}$. }
\label{fig:ExTh}
\includegraphics[scale=0.55]{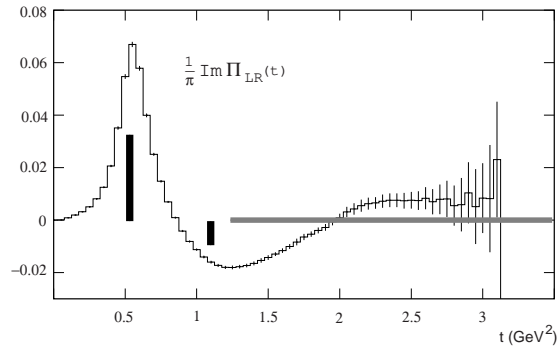}
\end{figure}

The spectral function of the correlation function defined in
Eq.~\rf{LR} can be obtained from measurements of the hadronic
$\tau$--decay spectrum (vector--like decays minus axial--vector like
decays). We plot in Figure~\ref{fig:ExTh} the experimental
determination of
$\frac{1}{\pi}\Imm\Pi_{LR}(t)$, obtained from the ALEPH
collaboration data~\cite{ALEPH}, versus the invariant hadronic mass
squared $t$ in the accessible region $0\le t\le m_{\tau}^2$.
In this case, the corresponding spectrum of the $\MHA$ to
$\QCD_{\infty}$ consists of the pion pole (not shown in the
figure), a vector narrow state(the $\rho$)  and an axial--vector
narrow state (the $A_1$). At this level of approximation, and in the
chiral limit, the rest of the vector and axial--vector states
are taken to be degenerate and  cancel in the difference, a fact which is
simulated by the horizontal continuum line in the figure.  Looking
at this plot, one can hardly claim that this approximation reproduces
the details of the experimental data. \underline{However}, with
$\Pi_{LR}(Q^2)$ determined from the spectral function by the unsubtracted
dispersion relation
\be
\Pi_{LR}(Q^2)=\int_{0}^{\infty}dt\frac{1}{t+Q^2}
\frac{1}{\pi}\Imm\Pi_{LR}(t)\,,
\ee 
the corresponding plot of the function
$\frac{-Q^2}{F_{0}^2}
\Pi_{LR}(Q^2)$ versus the euclidean variable $Q^2=-t$ is shown in
Fig~\ref{fig:EuLRAA}. The solid curve is the one corresponding to the
simple
$\MHA$ in Eq.~\rf{MHA} and the dotted curve the one from the experimental
data in Fig~\ref{fig:ExTh}. One can see that, by contrast to what happens
in the Minkowski region shown in Fig.~\ref{fig:ExTh}, the corresponding
curves in the euclidean region are both very smooth and in fact the
$\MHA$ already provides a rather good interpolation between the asymptotic
regimes where, by construction, it has been constrained to satisfy the
lowest order chiral behaviour in Eq.~\rf{chpt}, and the two Weinberg sum
rules: the OPE constraints in Eqs.~\rf{1wsr} and
\rf{2wsr}. This  good interpolation is the reason why the integral in
Eq.~\rf{pimd}, evaluated at the $\MHA$, already reproduces the experimental
result rather well .

\subsection{Other Analytic
Approaches}

\begin{figure}
\caption{\it Plot of $\frac{-Q^2}{F_{0}^2}
\Pi_{LR}(Q^2)$ in the euclidean region. The
solid curve is the one corresponding to the simple $\MHA$ in
Eq.~\rf{MHA} and the dotted curve the one from the experimental data
in Fig~\ref{fig:ExTh}. The other curves are the predictions of various
models discussed in the text.}
\label{fig:EuLRAA}
\includegraphics[scale=0.65]{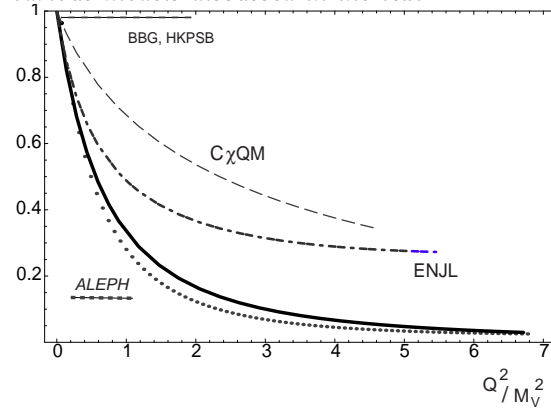}
\end{figure}

At this stage, it is illustrative to compare the large--$N_c$
approach we have discussed so far with other analytic
approaches in the literature. The $\Pi_{LR}$
correlation function provides us with an excellent theoretical 
laboratory to do this comparison. The different shapes of
$\frac{-Q^2}{F_{0}^2}
\Pi_{LR}(Q^2)$ predicted by other analytic approaches are collected in
Fig.~\ref{fig:EuLRAA}.
Here, several comments are in order.

a) The suggestion to use a large--$N_c$ QCD framework combined with 
$\chi$PT cut--off loops, was first proposed by Bardeen, Buras and G\'erard
in a series of seminal papers~\cite{Bu88,Ba89,Ge90}. The same method has
been applied by the Dortmund group~\cite{Dortmmund}, in particular to
the evaluation of
$\epsilon'/\epsilon$. In this approach the {\it hadronic ansatz}
to the Green's functions consists of Goldstone poles \underline{alone}
and, therefore, integrals like the one in Eq.~\rf{pimd}  become
UV--divergent (often quadratically divergent) since the correct QCD
short--distance behaviour is not implemented. In practice the integrals 
are cut at a {\it physical} cut off: $\Lambda\sim 1~\GeV$. In the case we
are considering, the predicted shape of the LR--correlation function
normalized to its value at
$Q^2=0$ is a constant all the way up to the chosen cut--off value, as shown
by the BBG, HKPSB line  (the horizontal dotted
line) in Fig.~\ref{fig:EuLRAA}.

b) The Trieste group  evaluate the relevant Green's functions using
the constituent chiral quark model (C$\chi$QM) proposed in
refs.~\cite{MG84} and \cite{EdeRT90,PdeR91}. 
They have obtained a long list
of  predictions~\cite{Trieste}, in
particular
$\epsilon'/\epsilon$. The model gives an educated first guess of
the low--$Q^2$ behaviour of the Green's functions, as one can judge
from the C$\chi$QM--curve (dashed curve) in Fig.~\ref{fig:EuLRAA},
but it  fails to reproduce the short--distance QCD--behaviour. Another
objection to this approach is that the ``natural matching scale'' to
the short--distance behaviour in this model should be
$\sim 4M_{Q}^2$ ($M_{Q}$ the constituent quark mass), which is too low to
be trusted.

c) The extended Nambu--Jona-Lasinio (ENJL) model was developed as an
improvement on the C$\chi$QM, since in a certain way it incorporates
the vector--like fluctuations of the underlying QCD theory which are
known to be phenomenologically important (see
e.g. refs.~\cite{BBdeR93,Bij96} where other references can also be found). 
The model is rather successful in predicting the low--energy
$O(p^4)$ constants of the chiral Lagrangian~\cite{BBdeR93}. It has, indeed,
a better low--energy behaviour than the C$\chi$QM, as the ENJL--curve
(dot--dashed) in Fig.~\ref{fig:EuLRAA} shows; but it fails to
reproduce the short--distance behaviour of the OPE in QCD. 
Arguments, however, to do
the matching to short--distance QCD have been forcefully  elaborated in
refs.~\cite{LuGr}, which also give a lot of numerical predictions; a
large value for
$\epsilon'/\epsilon$ in particular.

The problem with the ENJL
model as a plausible model of large--$N_c$ QCD, is that the on--shell
production of unconfined constituent quark $Q\bar{Q}$ pairs that it
predicts, violates the  large--$N_c$ QCD counting rules.
In fact, as shown in ref.~\cite{PPdeR98}, when the unconfining pieces
in the ENJL spectral functions are removed by adding an appropriate
series of local counterterms, the resulting theory is entirely
equivalent to an effective chiral meson theory with three narrow
states V, A and S; very similar to the phenomenological {\it Resonance
Chiral Lagrangians} proposed in refs.~\cite{EGPdeR89,EGLPdeR89}.
These Lagrangians can be viewed, therefore, as particular models
of large--$N_c$ QCD. They predict the same  Green's functions as the
$\MHA$ to $\QCD_{\infty}$ discussed above, in 
some particular cases but not in general (see e.g. the
three--point functions discussed in ref.~\cite{KN01}).  
  

In view of the difficulties which the analytic approaches discussed in a),
b) and c) above, have in reproducing the shape of  the simplest Green's
function one can think of, it is difficult to attribute more than a 
qualitative significance to their ``predictions'';
$\epsilon'/\epsilon$ in particular, which requires the interplay of
several other Green's functions much more complex than
$\Pi_{LR}(Q^2)$.

\section{ METHODOLOGY, APPLICATIONS}\label{section:MET}

The $\QCD_{\infty}$ approach that we are proposing in order to compute a
specific coupling of the chiral electroweak Lagrangian consists of the
following steps:

\begin{enumerate}
\item{\it Identify the relevant Green's functions.}

In
most cases of interest, the underlying QCD Green's functions in question
are two--point functions with additional zero momentum insertions of
vector, axial vector, scalar and pseudoscalar currents. The higher the
power in the chiral expansion, the higher will be the number of zero
momentum  insertions. This step is totally general and does not invoke any
large--$N_c$ approximation. 

\item{\it Work out the short--distance behaviour and the
long--distance behaviour of the relevant Green's functions.}

The long--distance behaviour is governed by the Goldstone
singularities and can be obtained from $\chi$PT. The
short--distance behaviour is governed by the OPE of the currents
through which the hard momenta flows. Again, this step is well
defined independently of the large--$N_c$ expansion; in practice,
however, the calculations simplify when restricted to the
appropriate order in the $1/N_c$--expansion one is interested in. In
particular, chiral loops are subleading in the $1/N_c$--expansion.

\item{\it Introduce a large--$N_c$ approximation for the underlying
Green's functions.}

As already mentioned, Green's functions in $\QCD_{\infty}$ are
meromorphic functions. The {\it minimal hadronic approximation} (MHA) that
we are proposing consists in limiting the number of poles to the minimum
required to satisfy the  leading power fall--off at
short--distances, as well as the appropriate
$\chi$PT long--distance constraints.  
\end{enumerate}

\noi
The three steps above can be done analytically, which helps to unravel the
intricacies of the underlying dynamics. The method is, in principle,
\underline{improvable} (unlike the other models
discussed above). It can be improved by adding more constraints from the
next--to--leading short--distance inverse power behaviour in the OPE and/or
higher orders in the chiral expansion. This has been tested within a toy
model of $QCD_{\infty}$ in ref.~\cite{GPdeR02}

We have checked this approach with the calculation of a few
low--energy observables:

{\em i) The electroweak $\Delta m_{\pi}$ mass
difference} which we have already discussed~\cite{KPdeR98}. 

{\em ii) The hadronic vacuum polarization contribution to the
anomalous magnetic moment of the muon
$a_{\mu}$}.
The MHA in this case requires one
vector--state pole and a perturbative QCD continuum. The absence of
dimension two operators in QCD in the chiral limit, constrains the
threshold of the continuum. The result thus obtained~\cite{PdeR} is
compatible with the more precise phenomenological determinations which use
experimental input.

{\em iii) The $\pi^{0}\ra e^{+}e^{-}$ and $\eta\ra \mu^{+}\mu^{-}$ decay
rates}. These processes are governed by a $\langle
PVV\rangle$ three--point function, with the $Q^2$--momentum flowing through
the two $V$--currents. The MHA in this case requires a vector--pole
and a  double vector--pole. The predictions~\cite{KPPdeR99} of the
branching ratios
$\frac{\Gamma(P\ra l^{+}l^{-})}{\Gamma(P\ra
\gamma\gamma)}$
are in good agreement with the present experimental determinations.

These successful predictions have encouraged some of us to pursue  a
systematic analysis of $K$--physics observables within the same
large--$N_c$ framework. So far, the following calculations have been made:

\begin{itemize}

\item{\it The $B_{K}$--Factor in the Chiral Limit}~\cite{PdeR00}.
\vspace*{-0.15cm}
\item{\it Weak Matrix Elements of the Electroweak Penguin Operators
$Q_7$ and $Q_8$}~\cite{KPdeR99,KPdeR01}.
\vspace*{-0.15cm}
\item{\it Hadronic Light--by--Light Scattering Contribution to the Muon
$g-2$}~\cite{KN02,KNPdeR02}.
\vspace*{-0.15cm}
\item{\it Electroweak Hadronic Contributions to the Muon
$g-2$}~\cite{KPPdeR02}.  

\end{itemize}

We also have preliminary results on:

\begin{itemize}

\item {\it The Long--Distance Contribution to the
$K_{L}\ra\mu^{+}\mu^{-}$ Decay Rate}~\cite{GdeR02}.
\vspace*{-0.15cm}
\item {\it Weak Matrix Elements of the Strong Penguin
Operators
$Q_4$ and $Q_6$}~\cite{HPdeR02}.

\end{itemize}

\section{FACTOR $B_{K}$ OF $K^{0}-\bar{K}^{0}$ MIXING}

The factor in question is conventionally defined by the matrix
element of the four--quark operator $Q_{\Delta S=2}(x)=
(\bar{s}_{L}\gamma^{\mu}d_{L})
(\bar{s}_{L}\gamma_{\mu}d_{L})(x)$:
\be
\langle \bar{K}^{0}\vert Q_{\Delta S=2}
(0)\vert
K^{0}\rangle =\frac{4}{3}f_{K}^2 M_{K}^2 B_{K}(\mu)\,.
\ee
To lowest order in the chiral expansion the operator  $Q_{\Delta
S=2}(x)$ bosonizes into an $O(p^2)$ term:
\be
-\frac{F_{0}^4}{4}g_{\Delta S=2}(\mu)
\underbrace{\left[(D^{\mu}U^{\dagger})U\right]_{23}
\left[(D_{\mu}U^{\dagger})U\right]_{23}}_{\frac{2}{F_{0}^2}
\partial^{\mu}K^{0}\partial_{\mu}K^{0}-(r^{\mu})_{23}
(r_{\mu})_{23}+\cdots}\,,
\ee
with $g_{\Delta S=2}(\mu)$ the low energy constant (which depends on the
renormalization scale $\mu$) to be determined. A convenient choice
of the underlying Green's function here is the four--point
function
$W_{LRLR}^{\mu\alpha\nu\beta}(q,l)$ of two left--currents
$(\bar{s}_{L}\gamma^{\mu}d_{L})$ and $(\bar{s}_{L}\gamma^{\nu}d_{L})$, 
which carry the
$q$--momentum one has to integrate over, and two right--currents
$(\bar{d}_{R}\gamma^{\alpha}s_{R})$ and
$(\bar{d}_{R}\gamma^{\beta}s_{R})$ with soft $l$--momentum insertions,
which couple to the external $(r_{\alpha})_{32}$-- and
$(r_{\beta})_{32}$--sources. The coupling constant
$g_{\Delta S=2}(\mu)$, which has to be evaluated in the same
renormalization scheme as the Wilson coefficient $C_{\Delta S=2}(\mu)$ has
been evaluated, is then given by the integral~\cite{PdeR00}
$$
g_{\Delta S=2}(\mu)\!=\!1-\frac{1}{32\pi^2
F_{0}^2}\!\left[\frac{\left(4\pi\mu^2\right)^{\epsilon/2}}
{\Gamma(2-\epsilon/2)}
\!\int_{0}^{\infty}\!\!\!\frac{dQ^2}{(Q^2)^{\epsilon/2}}\times\right.
$$
\be\lbl{bkint}
\left.\underbrace{\left(\frac{-g_{\alpha\beta}}{3}\right)
\!\!\int
\!d\Omega_{q}\lim_{l\ra
0}g_{\mu\nu}W_{LRLR}^{\mu\alpha\nu\beta}(q,l)}_{W_{LL}(Q^2)}\right]
_{\overline{MS}}\!\!\,,
\ee

\noi 
conceptually similar to the one which determines the electroweak
constant
$C$ in Eq.~\rf{pimd}. 
The renormalization--scale invariant
$\hat{B}_{K}$--factor is then given by the product
\be
\hat{B}_{K}=\frac{3}{4}C_{\Delta S=2}(\mu)\times g_{\Delta S=2}(\mu)\,.
\ee

The large--$N_c$ hadronic approximation of the Green's
function
$W_{LL}(Q^2)$ in Eq.~\rf{bkint},  which fulfills the leading OPE
short--distance constraint and the long--distance constraints which fix 
$W_{LL}(0)$ and
$W_{LL}^{`}(0)$ in $\chi$PT, requires one vector--pole, a double
vector--pole and a triple vector--pole~\footnote{In fact, this goes beyond
the strict MHA which here  only requires a vector--pole. It is
the {\it extra} input of $W_{LL}(0)$
and
$W_{LL}'(0)$, as known from $\chi$PT, which allows us to improve on
the strict MHA.} The integral in Eq.~\rf{bkint} can then be evaluated, with
the result~\cite{PdeR00}
\be\lbl{bk}
\hat{B}_{K}=0.38\pm 0.11\,.
\ee
The $\mu$--scale and renormalization scheme dependence in $g_{\Delta
S=2}(\mu)$ cancels with the one in $C_{\Delta S=2}(\mu)$, when evaluated
at the same approximation in the $I/N_c$--expansion. 

When comparing this
result to other determinations, specially in Lattice QCD, it should be
realized that the unfactorized  contribution in Eq.~\rf{bkint} is the one
in the chiral limit. It is possible, in principle, to calculate chiral
corrections within the same large--$N_c$ approach, but this has not yet
been done. 

The result in Eq.~\rf{bk} is compatible with 
the old current algebra prediction~\cite{DGH82} which, 
to lowest order in $\chi$PT, relates the
$B_{K}$--factor to the $K^{+}\ra \pi^{+}\pi^{0}$ decay rate. 
In fact, our
calculation can be viewed as a 
successful prediction of
the $K^{+}\ra \pi^{+}\pi^{0}$ decay rate~!

As discussed in ref.~\cite{PdeR96}, 
the bosonization of the four--quark operator
$Q_{\Delta S=2}$ and the bosonization of the operator
$Q_{2}-Q_{1}$ which generates $\Delta I=1/2$ transitions,   
are related to each
other in the combined chiral limit and  next--to--leading 
order in the $1/N_c$--expansion, when mixing with the penguin operators is
neglected. Lowering the value of
$\hat{B}_{K}$ from the factorized $\cO(N_c^2)$ prediction:
$\hat{B}_{K}=3/4$, to the result in  Eq.~\rf{bk}, is
correlated with an increase of the coupling constant $g_{\underline{8}}$ 
in the lowest order
effective chiral Lagrangian (see Eq.~\rf{g8}) which generates 
$\Delta I=1/2$ transitions, and
provides a first step towards a quantitative understanding of 
underlying dynamics of the  $\Delta I=1/2$ rule. 

\section{ELECTROWEAK PENGUINS}
We shall be next concerned with the four--quark operators generated by the
so called electroweak penguin diagrams of the Electroweak Theory
\be
\cL\Ra \cdots
C_{7}(\mu)Q_{7}+
C_{8}(\mu)
Q_{8}\,,
\ee
with $C_{7}(\mu)$, $C_{8}(\mu)$ the Wilson coefficients of the operators
\be
Q_7   =   6(\bar{s}_{L}\gamma^{\mu}d_{L})
\sum_{q=u,d,s} e_{q} (\bar{q}_{R}\gamma_{\mu}q_{R})\,, 
\ee
\be
Q_8   =  
-12\sum_{q=u,d,s}e_{q}(\bar{s}_{L}q_{R})(\bar{q}_{R}d_{L})\,.
\ee
These operators generate terms of $O(p^0)$ in the effective chiral
Lagrangian\cite{BW84}; therefore, their matrix elements, although
suppressed by an $e^2$ factor, are chirally enhanced. Furthermore, the
Wilson coefficient $C_{8}$ has a large imaginary part induced by the
top--quark integration, which makes the matrix elements of the
$Q_{8}$ operator to be particularly important in the evaluation of
$\epsilon'/\epsilon$. 

Within the large--$N_c$ framework, the bosonization of these
operators produce matrix elements with the following counting
\be\lbl{Q7Q8z}
\langle Q_{7}\rangle\vert_{O(p^0)}= \underline{O(N_c)}      
+O(N_c^0)\,,
\ee
\be
\langle Q_{8}\rangle\vert_{O(p^0)}= \underline{O(N_{c}^2)}
+\!\!\!\!\!\!\!\!\!
\underbrace{O(N_c^0)}_{{\mbox{\rm
Zweig suppressed}}}\,. 
\ee

\subsection{Bosonization of $Q_7$}
The
bosonization of the
$Q_{7}$ operator to
$O(p^0)$ in the chiral expansion and to $O(N_c)$ is very similar to
the calculation of the $Z$--contribution to the coupling constant $C$
in Eq.~\rf{pimd}. An evaluation which also takes into account the
renormalization scheme dependence has been recently made in
ref.~\cite{KPdeR01} with the result: 
$$
\langle Q_{7}\rangle\vert_{\cO(p^0)}  =  6\ 
\underbrace{\langle
0\vert
(\bar{s}_{L}\gamma^{\mu}d_{L})(\bar{d}_{R}\gamma_{\mu}s_{R})\vert 0 
\rangle}_{\langle O_{1}(\mu)\rangle}\times \nn
$$
\be
\tr\left( U\lambda_{L}^{(23)} U^{\dagger}
 Q_{R}\right)^{\dag}\,,\quad\annd\quad
\lambda_{L}^{(23)}=\delta_{i2}\delta_{j3}\,.
\ee 
Here, the vev $\langle O_{1}(\mu)\rangle$ is given by the integral
$$
\langle
O_{1}(\mu)\rangle=\left\{\frac{3(\epsilon
-3)}{16\pi^2}\frac{(4\pi\mu^2)^{\,
\epsilon/2}} {\Gamma(2-\epsilon/2)}\times\right.
$$
\be\int_{0}^{\infty}dQ^2
\left.(Q^2)^{1-\epsilon/2}\left(
-Q^2\Pi_{LR}(Q^2)\right)\right\}_{\overline{MS}}\,, 
\ee
with $\Pi_{LR}(Q^2)$ the \underline{same} correlation function as in
Eq.~\rf{LR}.  In $\QCD_{\infty}$ this integral can be, formally, evaluated
exactly. When restricted to the same MHA which has been used to calculate
$\Delta m_{\pi}$, one gets the simple result
\be\lbl{LRgralmsb}
\langle O_1\rangle =   
 \frac{3}{32\pi^2}\!\left[
f_{V}^2
M_{V}^{6}\ln\!\frac{\Lambda^2}{M_{V}^2}\!-\!f_{A}^2
M_{A}^{6}\ln\!\frac{\Lambda^2}{M_{A}^2}\right]\,,
\ee
where $\Lambda^2\!\!=\!\!\mu^2\exp(1/3+\kappa)$, with $\kappa\!\!=\!-1/2$
in the  naive dimensional renormalization scheme (NDR) and 
$\kappa\!\!=\!\!+3/2$ in the 't Hooft--Veltman scheme (HV).

\subsection{Bosonization of $Q_8$}
To lowest  $\cO(p^0)$ in the chiral expansion, the
four--quark operator $Q_8$ bosonizes as follows
\be
\langle Q_{8}\rangle\vert_{\cO(p^0)} \!=\! -12\hspace*{-0.6cm}
\underbrace{\langle
O_{2}(\mu)\rangle}_{
\langle 0\vert (\bar{s}_{L}s_{R})(\bar{d}_{R}d_{L})\vert
0\rangle}\hspace*{-0.6cm}
\tr\left( U\lambda_{L}^{(23)} U^{\dagger}
 Q_{R}\right)^{\dag}\!\!\,. \nn
\ee
As noted in refs.~\cite{DG00,KPdeR01}, the vev $O_{2}(\mu)$ also appears
in the Wilson coefficient of the
$1/Q^6$ term in the OPE of the \underline{same} 
$\Pi_{LR}(Q^2)$ correlation function as in Eq.~\rf{LR}, for which the
MHA to $\QCD_{\infty}$ gives a rather good approximation, as we have
already discussed. This offers the possibility of obtaining an estimate of
the vev $O_{2}(\mu)$ \underline{beyond} the large--$N_c$ approximation,
where 
$O_{2}\!\Ra\!\frac{1}{4}\stern^2$, and, therefore, without having to fix
a value for the
$\stern$--condensate, which is poorly known. Unfortunately, as one
can judge from the results in Table~\ref{table:1}, different methods to
extract the value of
$O_{2}(\mu)$ lead, at present, to rather different values for the weak
matrix elements:
\be\lbl{M78}
M_{7,8}\equiv\langle(\pi\pi)_{I=2}\vert Q_{7,8}\vert
0\rangle\,,\quad\mbox{\rm at}\quad \mu=2~\GeV\,.
\ee
In view of the diversity of results obtained with the dispersive
approach, depending on which type of sum rules are
used; and the unknown systematic errors of the quenched approximation and
the chiral limit extrapolations in the lattice results, it looks like it
will take some time before one can claim that these matrix elements are
known reliably.

\begin{table*}[htb]
\caption{Matrix Elements Results for $M_{7,8}$ (see Eq.~\rf{M78}) in
$\GeV^3$ using different methods.}
\label{table:1}
\newcommand{\cc}[1]{\multicolumn{1}{c}{#1}}
\renewcommand{\tabcolsep}{1pc} 
\renewcommand{\arraystretch}{1.4} 
\begin{tabular}{@{}lllll}
\hline \hline
METHOD  & $M_7$(NDR) & $M_7$(HV) & $M_8$(NDR) &
$M_8$(HV)\\
\hline
{\sc Large--$N_c$ $\MHA$} & & & &  
\\ {\it Knecht, Peris, de~Rafael} & $0.11\pm 0.03$ &
$0.67\pm 0.20$ & 
$2.3\pm 0.7$ & $2.5\pm 0.8$ \\
(with $\als^2$ corrects.)~\cite{KPdeR01,K01P01} & & &  &
\\
\hline 
{\sc Dispersive Approach} & & & & 
\\
{\it Narison}~\cite{Na01}  & $0.17\pm 0.05$
&  &
$1.4\pm 0.3$ & 
\\ {\it  Cirigliano et al}~\cite{CDGM01} & $0.16\pm 0.10$ & $0.49\pm
0.07$ & $2.2\pm 0.7$ & $2.5\pm 0.7$ 
\\
{\it Bijnens  et al}~\cite{BGP01} &
$0.24\pm 0.03$ &
$0.37\pm 0.08$ &
$1.2\pm 0.8$ & $1.3\pm 0.8$
\\
{\it  Cirigliano et al}~(OPAL data)~\cite{CDGM02} & $0.21\pm 0.05$ &  &
$1.7\pm 0.3$ &
\\
\hline 
{\sc Lattice QCD}  & & & & 
\\ {\it Bhattacharia et al}~\cite{Bhatetal} & $(0.32\pm 0.06)$
 &
 & $(1.2\pm 0.2)$ & 
\\{\it  Donini  et al}~\cite{Donetal} &
$0.11\pm 0.04$ &
$0.18\pm 0.06$ &$ 0.51\pm 0.10$ & 
$0.62\pm 0.12 $
\\
RBC {\it coll.}~\cite{RBC} & $(0.27\pm 0.03)$ & & $
(1.1\pm 0.2)$ &\\ CP-PACS {\it coll}~\cite{CPPACS}
& $(0.24\pm 0.03)$ & & $(1.0\pm 0.2)$ & 
 \\ \hline
\hline
\end{tabular}\\[2pt]
Only the results obtained
with methods which can exhibit an explicit dependence on the
renormalization scale are quoted in this Table. The numbers in brackets
have been obtained after informal private discussions with some of the
authors of the various lattice collaboration.  
\end{table*}

\subsection{Test of the Zweig Rule}
Besides the important issue of getting
an accurate determination of the
$Q_8$ matrix elements, a reliable determination of the vev $\langle
0\vert (\bar{s}_{L}s_{R})(\bar{d}_{R}d_{L})\vert 0\rangle$ would be most
welcome as a test of the Zweig rule in the scalar and pseudoscalar
sectors. More precisely,
\be
\langle
O_{2}(\mu)\rangle=\frac{1}{4}\stern^2+ 
\langle 0\vert (\bar{s}_{L}s_{R})(\bar{d}_{R}d_{L})\vert
0\rangle_{c}\,,
\ee
where the unfactorized contribution (the second term) involves Feynman
diagrams which require gluon exchanges between at least two quark loops.
These are  the so called Zweig--suppressed contributions, which are indeed
$\cO(N_c^0)$ in the $1/N_c$ expansion. 
There are reasons to
suspect that subleading terms in the $1/N_c$ expansion involving Green's
functions of scalar (and pseudoscalar) density operators might be
important, unlike those which only involve vector (and axial-vector)
currents. There are several phenomenological examples of this: the
$\eta'$ mass, the possible existence of a broad $\sigma$ meson, large final
state interactions in states with $J=0$ and $I=0$,
etc. We are considering the possibility that the
appropriate expansion for these exceptional Green's functions could be a 
$1/N_c$ expansion in which $n_{f}/N_c$ is held fixed, where $n_f$ denotes 
the number of
light flavours; the kind of  expansion originally 
advocated by G.~Veneziano~\cite{V76}.

Formally, the connected part of the  vev $
\langle 0\vert (\bar{s}_{L}s_{R})(\bar{d}_{R}d_{L})\vert
0\rangle_{c}$ is defined by the integral
\be\lbl{zweig}
\frac{1}{i}\left(\int\frac{d^D
q}{(2\pi)^D}\,
\Psi_{ds}(Q^2)\right)_{\overline{\mbox{\rm{\footnotesize
MS}}}}^{\mbox{\rm\tiny ren.}}\,,
\ee
with $\Psi_{ij}(Q^2)$ the two--point function 
$$
\frac{i}{4}\int d^4 x e^{iq.x}\left\{\langle 0\vert
T\left[\bar{d} d(x)\,\,
\bar{s} s(0)\right]\vert 0\rangle\right.-
$$
\be
\hspace*{2.3cm}\left.\langle 0\vert
T\left[\bar{d}\gamma_{5} d(x)\,\,
\bar{s}\gamma_{5} s(0)\right]\vert 0\rangle\right\}\,.
\ee
It would be very helpful to get some information on the vev $
\langle 0\vert (\bar{s}_{L}s_{R})(\bar{d}_{R}d_{L})\vert
0\rangle_{c}$ from lattice QCD.  Ultimately, this could provide a way to
focus on the origin of the discrepancies in Table~\ref{table:1}.

\section{GLUONIC PENGUINS}
Finally, I shall make a few  analytic remarks concerning the
sector of the four--quark operators generated by the strong
Penguins of the Electroweak Theory
\be
\cL\Ra \cdots
C_{4}(\mu)Q_{4}+
C_{6}(\mu)
Q_{6}\,,
\ee
with $C_{4}(\mu)$, $C_{6}(\mu)$ the Wilson coefficients of the operators
\be
Q_4   =   4\sum_{q=u,d,s}(\bar{s}_{L}\gamma^{\mu}q_{L})
(\bar{q}_{L}\gamma^{\mu}d_{L})\,, 
\ee
\be
Q_6   =  
-8\sum_{q=u,d,s}(\bar{s}_{L}q_{R})(\bar{q}_{R}d_{L})\,.
\ee
These operators generate contributions to the coupling constant
$g_{\underline{8}}$ of the
$\cO(p^2)$ chiral Lagrangian in Eq.~\rf{g8}. The Wilson coefficient
$C_{6}(\mu)$  has also an important imaginary part generated by the
integration of the heavy flavour degrees of freedom, which makes the
operator
$Q_6$ particularly relevant for the evaluation of
$\epsilon'/\epsilon$ in the Standard Model. 

Seen from the large--$N_c$
framework point of view, the only terms which have been retained (so
far) in analytic evaluations of weak matrix elements of four--quark
operators, is best illustrated by
showing the terms which, equivalently, would be retained in
the lowest order anomalous dimension matrix of these operators. Non--zero
entries in this matrix appear then in three blocks, in the sector of the
$Q_1$,
$Q_2$,
$Q_4$,
$Q_6$ and $Q_8$ operators,  as follows:
\bea\lbl{adm}
\left(
\begin{array}{ccccc}
0 & \frac{3}{N_c} &  &  &    \\
\frac{3}{N_c} & 0 &  &  &  
\\
   &  & \frac{1}{3}
\frac{n_f}{N_c} 
 & \frac{1}{3}\frac{n_f}{N_c}&   
\\
   &  & \frac{1}{3}\frac{n_f}{N_c}
 &
-3+\frac{1}{3}\frac{n_f}{N_c} &   
\\
    & 
&  &  &
-3+\frac{3}{N_{c}^2} 
\\
\end{array}
\right)
\eea
More precisely, the mixing between matrix elements of the
$(Q_{2},Q_{1})$--operators on the one hand and those of the penguin--like
operators on the other, is neglected. Furthermore, in the 
$(Q_{2},Q_{1})$--sector, only the leading and next--to--leading terms in
the $1/N_c$--expansion have been retained~\cite{PdeR96,PdeR00}. Mixing
between the strong penguin sector and the electroweak penguin sector is
also neglected, but terms
$\cO(\frac{n_f}{N_c})$ in the mixing of $Q_{4}$ and $Q_{6}$ are retained.
Exceptionally, for reasons already discussed above, the
Zweig suppressed $\cO(1/N_c^2)$--corrections in $Q_8$ are also retained. I
can now discuss some recent analytic observations which have been made.

\subsection{Final State Interactions and $\epsilon'/\epsilon$}

The importance of final state interactions in the evaluation of
$K\ra\pi\pi$ matrix elements has been reexamined in a series of
recent papers~\cite{PP01}. The main observation is that
a simple numerical estimate of $\epsilon'/\epsilon$ can be obtained if
one proceeds as follows:

i) The evolution of the relevant Wilson coefficients from the $M_W$--scale
to the $m_c$--scale is retained, exactly, to next to leading order in pQCD.

ii) The bosonization of the relevant $Q_6$ and $Q_8$ operators is done
at the leading $\cO(N_c^2)$ \underline{only}. Notice that this corresponds
to the {\it further} restriction where the matrix in Eq.~\rf{adm}
becomes diagonal, with only non--zero entries in the ($Q_6$, $Q_8$) sector:
$
\left(
\begin{array}{cc}
-3 & 0   
\\
0 &
-3
\\
\end{array}
\right)
$. Starting at this level, the only $1/N_c$ contributions which are then
retained, are those generated by  an
approximate estimate of the the $\pi\pi$ final state interactions.

iii) The $\pi\pi$ final state interactions are approximated by an
Omn\`es--like resummation of the leading $\pi\pi$ chiral loops with
non--zero discontinuities. The authors of ref.~\cite{PP01} justify this
procedure by the fact that it reproduces rather well some of the
phenomenologically known
$\cO(p^4)$ local terms of the electroweak chiral Lagrangian. 

The prediction thus obtained
\be
\Ree(\frac{\epsilon\prime}{\epsilon})\!=\!\left(\!1.7\pm\!
\underbrace{0.2}_{\mu-{\mbox{\rm\tiny
scale}}} \underbrace{\begin{array}{c} +0.8
\\ -0.5
\end{array}}_{\stern}\!\pm\! \underbrace{0.5}_{1/N_c}\right)\!\!\times\!\!
10^{-3}\,, 
\ee
where their estimates of the various sources of errors are indicated in
underbracing, agrees within errors with the present experimental world
average~\cite{WAee} from NA31, NA48 and KTeV:
\be
\Ree(\epsilon\prime/\epsilon)\big\vert_{\mbox{\rm\scriptsize Exp.
}}=\left(1.66\pm 0.16\right)\times 10^{-3}\,.
\ee
The obvious question here is to know how stable is this prediction when
terms of $\cO(N_c)$  are also retained in the bosonization of $Q_6$ and
$Q_8$. The scenario with large deviations from the naive $\cO(N_c^2)$
bosonization of $Q_6$ and $Q_8$, which cancel in their overall
contribution to $\epsilon'/\epsilon$, cannot be excluded.  Let us not
forget that the restriction to
$\cO(N_c^2)$ terms in the bosonization of the four--quark operators, fails
dramatically to reproduce the
$\Delta I=1/2$ rule; a fact which, perhaps, should not be so surprising
since, after all, it is only when the
$\cO(N_c)$ terms are incorporated as well, that the
$\QCD_{\infty}$--properties of {\it planarity} apply.

\subsection{Bosonization of $Q_6$}

Related to the previous discussion is the question of the
bosonization of the  $Q_6$--operator.  It has been known for sometime
that this operator, to $\cO(N_c^2)$, gives a contribution to the
coupling constant $g_{\underline{8}}$ in Eq.~\rf{g8} which is modulated by
the product of the ratio $\stern^2/F_{0}^6$ times the $\cO(p^4)$
$L_5$--coupling of the strong chiral Lagrangian (known from the
$f_{\pi}/f_{K}$ ratio). It can be shown that, in the presence of
unfactorized terms, this contribution is modified as
follows~\cite{HPdeR02}:
$$
g_{\underline{8}}\vert_{Q_6}\!=\!C_{6}(\mu)\left\{\!-16
L_{5}\frac{\stern^2}{F_{0}^6}\!\!+\!\!
\left[\frac{1}{2\pi^2
F_{0}^4}\!\frac{\left(4\pi\mu^2\right)^{\epsilon/2}}{\Gamma(2-\epsilon/2)}
\right.\right.
$$
\be\lbl{Q6}
\left.\left.
\hspace*{1cm}\times\int_{0}^{\infty}\!\!\!dQ^2(Q^2)^{1-\epsilon/2}
\cW_{DG}(Q^2)
\right]_{\overline{MS}}\right\}\,,
\ee
where the function $\cW_{DG}(Q^2)$ is the analog of the
function $W_{LL}(Q^2)$ in Eq.~\rf{bkint} and the function $\Pi_{LR}(Q^2)$
in Eqs.~\rf{pimd} and \rf{LR}. Here, $\cW_{DG}(Q^2)$ is the invariant
function associated with a  four--point function
$\cW_{DGRR}^{~~~~\alpha\beta}(q,l)$ of two density--currents
$(\bar{s}_{L}q_{R})$ and $(\bar{q}_{R}d_{L})$, 
which carry the
$q$--momentum one has to integrate over, and two right--currents
$(\bar{d}_{R}\gamma^{\alpha}u_{R})$ and
$(\bar{u}_{R}\gamma^{\beta}s_{R})$ with soft $l$--momentum insertions,
which couple to the external $(r_{\alpha})_{12}$-- and
$(r_{\beta})_{31}$--sources.  We have found that the effect of the
unfactorized piece in Eq.~\rf{Q6} is very important. The large
effect can already be seen in the low $Q^2$ behaviour of the
function $\cW_{DG}(Q^2)$ in $\chi$PT
($B\equiv\frac{\vert\stern\vert}{F_{0}^2})$:
$$
\lim_{Q^2\ra 0}\cW_{DG}(Q^2)=\frac{n_f}{8}\left(\frac{BF_0}{Q^2}
\right)^2
$$
\be\lbl{WDG}
\hspace*{2cm}-n_f\left(
L_5-\frac{5}{2}L_{3}\right)\frac{B^2}{Q^2}+\cdots\,,
\ee
where the factor $L_5-\frac{5}{2}L_{3}\simeq 11\times 10^{-3}$, which
governs the behaviour of $Q^2\cW_{DG}(Q^2)$ at the origin (the integral
of the first term vanishes in dimensional regularization),  is very large
as compared to the individual values of
$L_5$ and
$L_3$. This large value results mostly from the effect of the lowest vector
state which the first $\cO(N_c^2)$ term in Eq.~\rf{WDG} does not see.
Clearly, this has serious implications
\underline{both} for the
$\Delta I=1/2$ rule and $\epsilon'/\epsilon$. An evaluation, along these
lines, of the coupling $g_{\underline{8}}\vert_{Q_6}$ is in
progress. 
       
\section{CONCLUSIONS and OUTLOOK}
We think that large--$N_c$ QCD provides a very useful framework to
formulate approximate calculations of the low--energy constants of the
effective chiral Lagrangian, both in the strong and electroweak sector.

The approach that we propose has been tested successfully with the
calculation of a few low--energy observables discussed in
section~\ref{section:MET}. Here, we have described various applications
to the evaluation of weak matrix elements which, for the first time in
analytic calculations of this type, explicitly show the cancellation of the
renormalization scale between the short--distance contributions and the
long--distance contributions. The analytic implementation of the approach
is sufficiently simple so as to provide an explanation of the dominant
underlying physical effects. It may also help, as a guidance, to unravel
important physical effects in Lattice QCD numerical simulations. 

\vspace*{0.5cm}
\noi
{\bf Acknowledgements}

\noi
My knowledge on the subjects reported here owes much to work and
discussions with my collaborators Santi Peris and Marc Knecht; and also,
at different stages, with M.~Perrottet, A.~Pich, J.~Bijnens, M.~Golterman,
T.~Hambye, A.~Nyffeler, and my students: B.~Phily, S.~Friot and
D.~Greynat.  I take this opportunity to thank them all. Special
thanks to Laurent Lellouch and Santi Peris for a careful reading of the
manuscript.

\end{document}